\documentclass[aps,prl,showpacs,notitlepage,nofootinbib,superscriptaddress,floatfix,showkeys,twocolumn]{revtex4-1}
\usepackage{blindtext}
\usepackage{hyperref}
\usepackage{amsmath,amssymb}
\usepackage{float}
\usepackage{microtype}
\usepackage{graphicx}
\usepackage{bm}
\usepackage{latexsym}
\usepackage{epsfig}
\usepackage{psfrag}
\usepackage{color}
\usepackage[dvipsnames]{xcolor}
\usepackage{subfigure}
\usepackage[section]{placeins}
\usepackage{comment}
\def\e1i{\epsilon_{1\mathrm{i}}}

\allowdisplaybreaks[1]

\begin{document}
%%%%%%%%%%%%%%%%%%%%%%%%%%%%%%%%%%%%%%%%%%%%%%%%%%%%%%%%%%%%%%%%%%%%%%%%%%%%%%%

\title{
$\gamma$ rays from in-flight positron annihilation as a probe of new physics
}

\author{Pedro De la Torre Luque}\email{pedro.delatorre@uam.es}
\affiliation{Departamento de F\'{i}sica Te\'{o}rica, M-15, Universidad Aut\'{o}noma de Madrid, E-28049 Madrid, Spain}
\affiliation{Instituto de F\'{i}sica Te\'{o}rica UAM-CSIC, Universidad Aut\'{o}noma de Madrid, C/ Nicol\'{a}s Cabrera, 13-15, 28049 Madrid, Spain}
\affiliation{The Oskar Klein Centre, Department of Physics, Stockholm University, Stockholm 106 91, Sweden}
\author{Shyam Balaji}
\email{shyam.balaji@kcl.ac.uk}
\affiliation{Physics Department, King’s College London, Strand, London, WC2R 2LS, United Kingdom}

\author{Pierluca~Carenza}\email{pierluca.carenza@fysik.su.se}
\affiliation{The Oskar Klein Centre, Department of Physics, Stockholm University, Stockholm 106 91, Sweden}

\author{Leonardo Mastrototaro}
\email{lmastrototaro@unisa.it}
\affiliation{Dipartimento di Fisica ``E.R. Caianiello'', Università degli Studi di Salerno, Via Giovanni Paolo II, 132 - 84084 Fisciano (SA), Italy}
\affiliation{INFN - Gruppo Collegato di Salerno, Via Giovanni Paolo II, 132 - 84084 Fisciano (SA), Italy.}

\smallskip
\begin{abstract}
The $\gamma$ ray emission originating from in-flight annihilation (IA) of positrons is a powerful observable for constraining high-energy positron production from exotic sources. By comparing diffuse $\gamma$ ray observations of INTEGRAL, COMPTEL and EGRET to theoretical predictions, we set the most stringent constraints on electrophilic feebly interacting particles (FIPs), thereby proving IA as a valuable probe of new physics. In particular, we extensively discuss the case of MeV-scale sterile neutrinos, where IA sets the most stringent constraints, excluding  $|U_{\mu4}|^{2} \gtrsim 10^{-13}$ and $|U_{\tau4}|^{2} \gtrsim 2\times 10^{-13}$ for sterile neutrinos mixed with $\mu$ and $\tau$ neutrinos respectively. These constraints improve existing limits by more than an order of magnitude. We briefly discuss the application of these results to a host of exotic positron sources such as dark photons, axion-like particles, primordial black holes (PBHs) and sub-GeV dark matter (DM). 
\end{abstract}
\maketitle

\emph{Introduction---}A steady production of positrons in the Milky Way is evidenced by the long-standing observations of a diffuse $511$~keV $\gamma$ ray line from electron-positron annihilation, whose emission seems constant over time~\cite{Siegert:2015knp,Kierans:2019aqz}. A full understanding of the observed characteristics of this emission is still challenging, although these observations allow us to constrain the properties of a variety of Galactic positron emitters \cite{Prantzos:2010wi, kierans2019positron, Siegert_2023}.
As a direct consequence of this positron injection, interactions of positrons with the interstellar medium (ISM) produce not only a diffuse line signal but leads also to a continuum emission that extends above and below $511$~keV. Here we list the processes responsible for a $\gamma$ ray signal associated with high-energy positrons:
\begin{itemize}
\item relativistic positrons lose energy by scattering on free and bounded electrons in the ISM, leading to a continuum emission above $511$ keV (IA emission);
\item thermalized positrons produce a para-positronium (p-ps) bound state with free electrons $25\%$ of the time, which annihilates in $\sim0.12$~ns~\cite{Badertscher_2007} into two $\gamma$ rays with $511$~keV energy each
\item in $75\%$ of the cases, thermalized positrons produce an ortho-positronium (o-ps) bound state with free electrons, which decays in $\sim140$~ns into (predominantly) $3$ photons~\cite{Badertscher_2007} with energy lower than $511$~keV.
\end{itemize}
A variety of different positron sources are known, such as pulsar wind nebulae, X-ray binaries (XRB), $\beta^+$ decay of unstable elements synthesized in supernovae (SN)~\cite{Prantzos:2010wi}, and cosmic ray (CR) interactions with the interstellar gas~\cite{Gabici:2019jvz}. In addition, ``exotic'' sources might also contribute to the positron flux. For instance, feebly interacting particles (FIPs) produced in SN explosions, and decaying into electrons and positrons (such as sterile neutrinos, dark photons or axion-like particles)~\cite{DelaTorreLuque:2023nhh,DelaTorreLuque:2023huu}, low-mass evaporating primordial black holes (PBHs) formed in the early Universe or sub-GeV dark matter (DM)~\cite{Koechler:2023ual,DelaTorreLuque:2023cef,DelaTorreLuque:2023olp}. These exotic sources are expected to produce positrons in the MeV-GeV range and are often not very well constrained at high energies, either by direct CR measurements or from their secondary $\gamma$ ray emissions, such as via inverse-Compton (IC) or bremsstrahlung.

In this letter, we propose the use of diffuse $\gamma$ ray emission produced from IA to constrain the production of positrons from 
various exotic high-energy emitters, extending the idea presented in Ref.~\cite{Beacom:2005qv} to a positron injection with an arbitrary new physics source and energy spectrum. We will focus on electrophilic FIPs produced in SN explosions, extensively discussing the case of sterile neutrinos. We also remark that IA leads to very competitive constraints for evaporating light PBHs and sub-GeV DM as well.

\emph{Positron injection and propagation---} Positrons injected in the Galaxy will diffuse in the Milky Way and lose energy when interacting gas, the Galactic magnetic field, and interstellar radiation fields, during their propagation, leading to a secondary diffuse $\gamma$ ray flux. In order to evaluate the effect of positron diffusion, we follow a similar procedure as in Refs.~\cite{DelaTorreLuque:2023huu, DelaTorreLuque:2023olp}. The spatial distribution and energy spectra of the positrons are computed with a recent version~\cite{de_la_torre_luque_2023_10076728} of the {\tt DRAGON2} code~\cite{Evoli:2016xgn, Evoli:2017vim}, a dedicated CR propagation code designed to numerically solve the full diffusion-advection-reacceleration-loss equation for the transport of charged particles in the Galactic environment~\cite{Ginz&Syr}. 
The spatial distribution of positron sources depends on the specific type of emitter that we consider. In this letter, we will focus on FIPs produced in SN explosions and subsequently decaying into electron-positron pairs. In this case, the spatial distribution follows the SN one, taken to be the \textit{Ferriere} distribution~\cite{ferriere2007spatial}. 

The set of propagation parameters that we use, obtained through a combined fit of different CR observables, allows us to reproduce the AMS-02~\cite{AGUILAR20211, aguilar2018observation, AMS_gen} and Voyager-1~\cite{VoyagerMO, Stone150} data in the range of some tens of MeV to the TeV scale~\cite{DelaTorreLuque:2023olp}\footnote{The input and outputs files from the used propagation code are available at \url{https://doi.org/10.5281/zenodo.10076728}.}.

After injecting positrons in the Galaxy, they will reach a steady-state distribution, with positrons annihilating once they reach the thermal energy of a warm medium (T$\simeq 8000$~K), as suggested by the analysis of the width of the $511$~keV line~\cite{Knoedlseder_2005}. Therefore, the positron distribution determines the spatial features of the 511~keV line emission~\cite{DelaTorreLuque:2023cef}. The 511~keV flux, integrated over the line of sight, is evaluated as~\cite{Guessoum1991}
\begin{equation}
    \frac{d\phi_{\gamma}^{511}}{d\Omega}=2k_{ps} \int ds \,\frac{d\phi^{e^+}}{d\Omega} \cdot n_e \cdot \sigma^\textrm{ann}(E_\textrm{th})\,,
\label{eq:511Line}
\end{equation}
where $k_{ps}=1/4$ is the fraction of positronium decays contributing to the 511~keV line signal, $\frac{d\phi^{e^+}}{d\Omega} = \int \frac{d\phi^{e^+}}{dE d\Omega} dE $ is the energy-integrated flux of positrons, $\sigma^{\rm ann}$ is conservatively approximated as the annihilation cross section of positrons with electrons at rest~\cite{Dirac} and the factor $2$ accounts for the emission of two photons per positron annihilation. Here, $d\Omega=dl\, db \cos b$ is the solid angle element being $l$, $b$ and $s$ the Galactic longitude, latitude and distance $s$ from the Earth. 

Since we expect that the electron density follows a smooth distribution in the disk of the Galaxy, we estimate the profile of the 511~keV line emission as directly proportional to the spatial distribution of diffuse positrons, and set the electron density to $n_{e}\simeq 1$~cm$^{-3}$, and uniform. Given that the electron density is suppressed out of the Galactic plane, we apply a scaling relation to the 511~keV profile following the vertical distribution of free electron density in the Galaxy~\cite{DelaTorreLuque:2023cef}, following the NE2001 model~\cite{Cordes:2003ik, Cordes:2002wz}. We have tested that this approach leads to consistent results with previous evaluations~\cite{Calore:2021klc,Calore:2021lih,DelaTorreLuque:2023huu,DelaTorreLuque:2023nhh,Carenza:2023old}. 

The correct evaluation of this flux is of primary importance since the predicted IA emission is directly proportional to the 511~keV line intensity. 
% This is a very conservative choice because: we expect a minimum in the total electron-positron interaction cross section around this temperature (see Fig.~3 of Ref.~\cite{Guessoum:2005cb}), and Ref.~\cite{Guessoum:2005cb} showed that the radiative recombination with electrons (e$^+$ + e$^-$ $\rightarrow$ Ps + $\gamma$) should be significantly more important than the direct annihilation (up to an order of magnitude higher~\cite{1976ApJ...210..582C, Guessoum:2005cb}). Charge exchange reactions are dominant above $\sim10^4$~K. 
Indeed, the IA emission is evaluated as~\cite{Beacom:2005qv}
\begin{equation}
    \begin{split}
        & \frac{d\phi^\textrm{IA}}{d\Omega \,dE_{\gamma}} = \frac{d\phi^{511}}{d\Omega} \frac{n_H}{P(1-\frac{3}{4}f)} \times  \\  & \times \int^{E_\textrm{max}}_{E_{\gamma}} dE'\,\frac{1}{N_{\rm pos}} \frac{dN_{\rm pos}}{dE'} \int^{E'} _{m_e}P_{E'\rightarrow E}\, \frac{d\sigma}{dE_{\gamma}} \frac{dE}{|dE/dx|} \,,
    \end{split}
    \label{eq:IA}
\end{equation}
where $\frac{d\phi^{511}}{d\Omega}$ is the flux per solid angle of the 511~keV line in the SPI region and the integral on $E'$ accounts for the energy-distribution of injected positrons determined by $dN_{\rm pos}/dE'$, being bounded between the highest energy used in the positron propagation simulation, $E_{\rm max}=5$~GeV, to the minimum positron energy able to produce $\gamma$ rays with energy $E_\gamma$. The term $\frac{1}{N_{\rm pos}} \frac{dN_{\rm pos}}{dE'}$ represents the fraction of positrons at the energy $E'$ contributing to the 511~keV emission (when this fraction is $1$, we recover the case of monoenergetic positrons generating the total 511~keV flux). Eq.~\eqref{eq:IA} takes into account the energy spectrum of injected positrons and it reduces to the expression used in Ref.~\cite{Beacom:2005qv} in the case of a monoenergetic injection.
The second integral takes into account the energy loss for a positron of energy $E'$ to the final energy before annihilating; and the number density of hydrogen, $n_H$, counts the number of targets on which positrons scatter and lose energy. Here, we defined $P_{E'\to E}$ as the probability, per unit of energy, for a positron with initial energy $E'$ to produce a $\gamma$ ray in flight before reaching the energy $E$, similarly $P=P_{E'\to m_{e}}$ is the probability for a positron with initial energy $E'$ to produce a $\gamma$ ray before thermalizing, and $f=0.967\pm0.022$~\cite{Jean:2005af} is the fraction of positrons annihilating at rest compared to the ones annihilating in-flight. 
We notice that this expression eventually becomes independent of the gas density, $n_H$, since $|dE/dx|$ (taken to be the ionization energy losses from the DRAGON code, including H and He) is directly proportional to $n_H$. Therefore, this factor does not involve additional uncertainties in the calculation of the IA emission.
The term $\frac{d\sigma}{dE_{\gamma}}$ is taken from Eq.~$(6)$ of Ref.~\cite{Beacom:2005qv}, which is normalized to a $\gamma$ ray multiplicity of $2$ photons (i.e. two photons per positron annihilated). Typically, the higher the energy of the positron, the lower the probability of annihilating with an electron and, for a given positron energy, the probability of producing a $\gamma$ ray is highest at the positron energy, and non-zero below it.

On top of the positron-induced $\gamma$ ray continuum signals, we emphasize that the Galactic IC and bremsstrahlung emission from CR electrons are the dominant $\gamma$ ray production processes below energies of a few hundreds of MeV. For this background, we use the electron model from Refs.~\cite{delaTorreLuque:2022vhm, DelaTorreLuque:2023zyd}, which is optimized to reproduce the electron and positron emission at Earth location, as well as the local $\gamma$ ray emissivity, down to a few tens of MeV. We have checked that this model shows strong agreement with the data down to $10$~keV. Some experiments employ simple power-laws fitted to data to characterize this background, while the IA emission has a very different spectrum in energy, helping to distinguish it from the background with high sensitivity. 

\emph{Feebly interacting particles as positron sources---} Novel particles dubbed FIPs are attracting increasingly more interest~\cite{Lanfranchi:2020crw,Agrawal:2021dbo,Alekhin:2015byh,Antel:2023hkf}. The most popular FIP models are axions~\cite{Raffelt:1987yt,Keil:1996ju,Carenza:2019pxu,Carenza:2020cis} and axion-like particles~\cite{Caputo:2021rux,Lella:2022uwi,Lella:2023bfb}, light CP-even scalars~\cite{Balaji:2022noj,Balaji:2023nbn,Dev:2021kje,Dev:2020jkh,Dev:2020eam}, sterile neutrinos~\cite{Kolb:1996pa,Raffelt:2011nc,Mastrototaro:2019vug} and dark photons~\cite{Chang:2016ntp}. 

Sub-GeV FIPs may be produced in copious amounts during a SN explosion, the final stage in the evolution of massive stars, thanks to the extreme temperature and density conditions reached in this astrophysical environment. After escaping the SN, FIPs might produce an energetic positron flux through their decay. This scenario was considered in previous works~\cite{DelaTorreLuque:2023nhh,DelaTorreLuque:2023huu}, considering the observational consequences of both the primary $e^\pm$ fluxes and the secondary photon fluxes. In this letter, we will explore this possibility further, by considering the IA emission of FIP-sourced positrons. Their injection spectrum is modeled with a modified blackbody spectrum~\cite{DelaTorreLuque:2023huu}.

Once the positron injection spectrum is determined, we derive the $95\%$ confidence level bounds by comparing theoretical predictions to the available diffuse $\gamma$ ray data in the MeV region, using mainly SPI~\cite{Bouchet:2008rp, Siegert:2015knp, Siegert:2022jii} (aboard INTEGRAL), COMPTEL~\cite{1998PhDT.........3K} and EGRET~\cite{Strong_2004} (both, aboard CGRO). We evaluate the bounds by calculating the best-fit value on the parameter that we constrain and the $2\sigma$ statistical error is associated through a $\chi^2$ test (in particular, we use the \verb|curve_fit| function in the \verb|scipy.optimize| Python package).

\begin{figure}[t!]
\includegraphics[width=\linewidth, height=0.21\textheight]{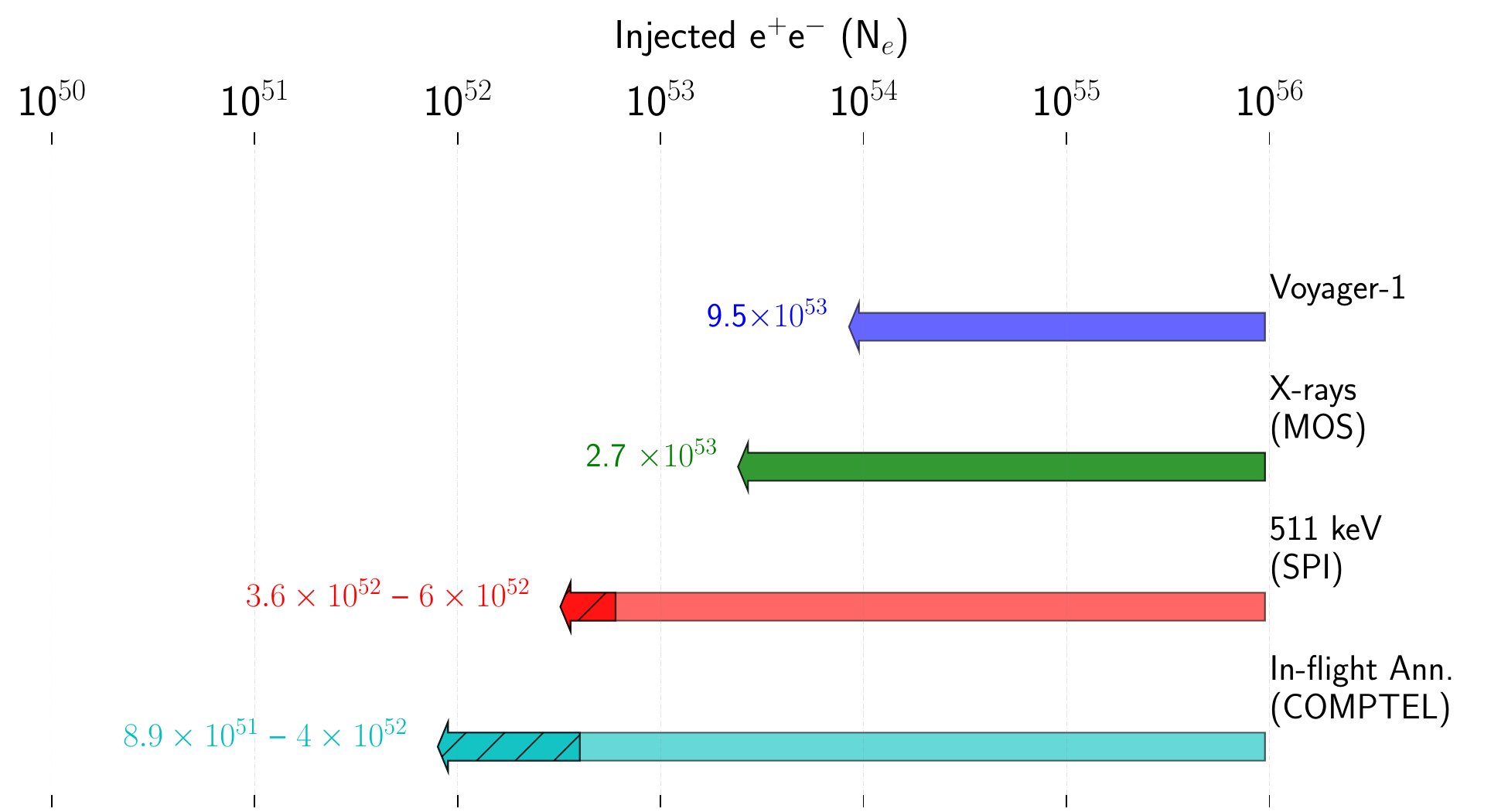}
\caption{Bounds at the $2\sigma$ level on the number of electrons and positrons injected per SN ($N_e$) for the general FIP case. We show here the limits derived from the local electron-positron flux by Voyager-1 (blue), X-ray emission from IC scattering of FIP-induced electron fluxes (green), the longitude profile of the $511$~keV line (red) and the ones derived comparing IA emission with the diffuse $\gamma$ ray observations of COMPTEL in the $|l|<30^{\circ}$ - $|b|<15^{\circ}$ region (cyan). The hatched bars represent the estimated uncertainties, as discussed in the text.}
\label{fig:FIP_Lim}
\end{figure}
In order to show the power of this observable in constraining new physics, we consider the generic FIP case, where we limit the number of $e^+$ injected by FIP decays per SN explosion, $N_{e}$. 
In Fig.~\ref{fig:FIP_Lim} we show the bounds on $N_{e}$ for the FIP case. These results lead to limits that are stronger than those previously found by considering the $e^\pm$ interstellar flux measured by Voyager-1~\cite{VoyagerMO}, secondary X-ray emission in the MOS detector energy range~\cite{Foster} and the 511~keV signal~\cite{DelaTorreLuque:2023nhh, DelaTorreLuque:2023huu}. Notice that we have updated the bound from the longitudinal profile of the 511~keV line emission (SPI data~\cite{Siegert:2015knp}) using the more accurate procedure detailed above, in very good agreement with the literature~\cite{Calore:2021lih,DelaTorreLuque:2023huu}. As discussed in Ref.~\cite{DelaTorreLuque:2023huu}, the bound from the longitude profile of the 511~keV line is severely affected by the systematic uncertainties at high longitudes, where the most constraining data points lie. As a proxy to evaluate the impact of systematic uncertainties at high longitudes, we repeat the analysis of the longitude profile considering only data points below $20^{\circ}$, which leads to constraint weaker by a factor $\simeq 2$ (hatched red region).
Finally, we show the bound obtained from IA emission using COMPTEL data in the $|l|<30^{\circ}$ - $|b|<15^{\circ}$ region~\cite{COMPTEL1994}. We have explored other regions of the sky, finding weaker constraints. In addition, we also add an uncertainty band represented by the cyan hatched region, showing the improvement in the bound when considering the background model as well. As we see, the bound on $N_e$ becomes between a factor of a few, to more than one order of magnitude better than the $511$~keV line constraint. Moreover, this bound could be improved even further by optimizing the region of interest as well as doing a more dedicated analysis of the raw data. This demonstrates that IA emission leads to the strongest constraints so far on electrophilic FIPs.

\begin{figure}[t!]
\includegraphics[width=0.99\linewidth]{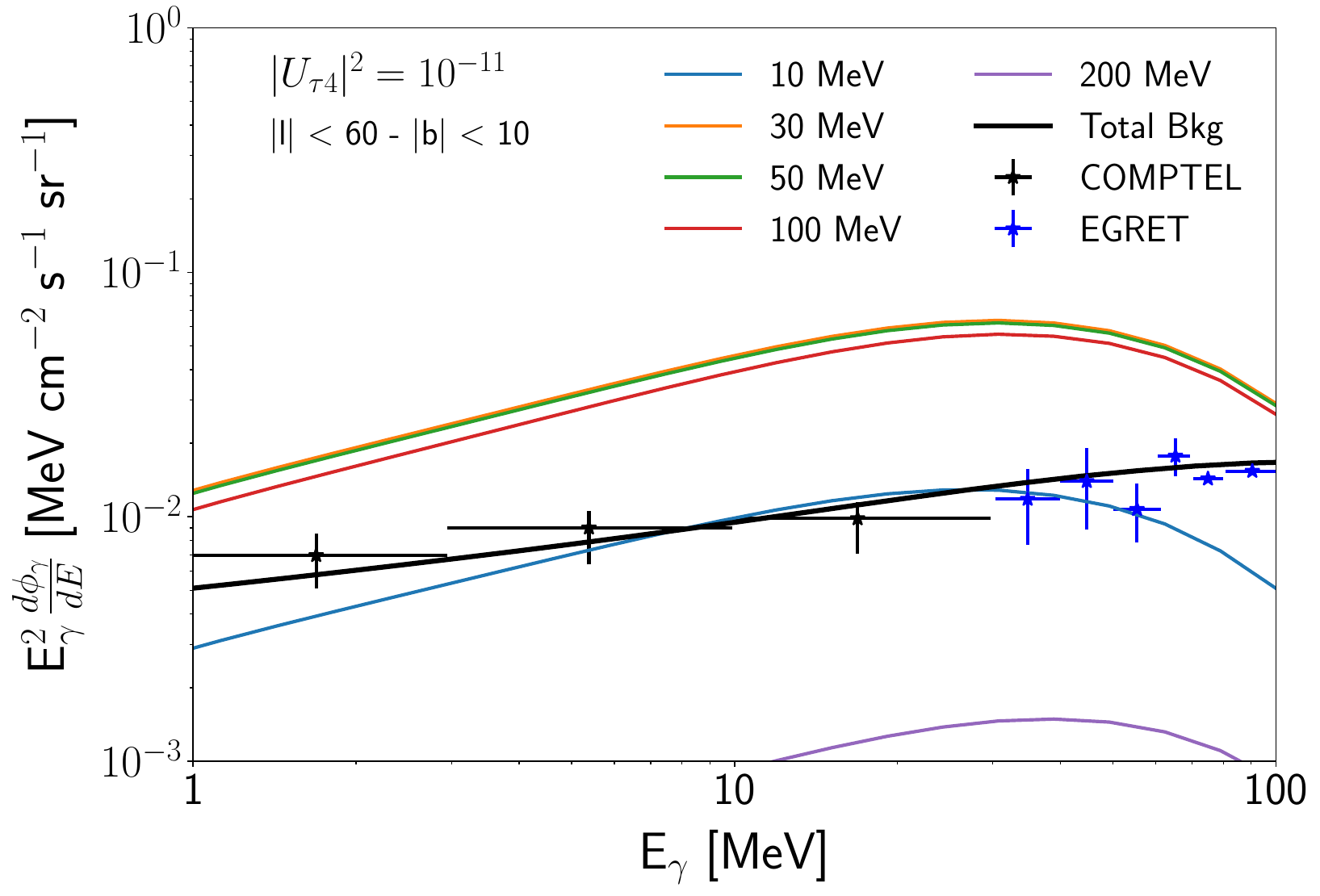}
\caption{Predicted IA spectra for $|U_{ \tau 4}|^2 = 10^{-11}$ and masses between $10$ and $100$~MeV, in the $|l|<60^{\circ}$ - $|b|<10^{\circ}$ region, compared to the COMPTEL and EGRET diffuse data.}
\label{fig:Snu_Main}
\end{figure}

\emph{The case of sterile neutrinos---} These results suggest that, for any FIP produced in a SN and decaying into positrons, the IA is an excellent probe. In the following, we consider the specific case of sterile neutrinos. In particular, we will focus on a MeV-GeV scale sterile neutrino, $\nu_4$ (indicating the mass eigenstate), mostly a flavour-sterile one, $\nu_s$ (indicating a flavor eigenstate), with a small mixing to $\mu$ or $\tau$ neutrinos $\nu_\alpha$, $\alpha=\mu,\,\tau$.
Sterile neutrinos have often been introduced to explain the origin of neutrino masses~\cite{Merle:2017dhf,Boyarsky:2018tvu,Abazajian:2012ys,Abazajian:2019ejt}, making them a well-motivated FIP. The positron spectrum injected by sterile neutrino decays is calculated by following Ref.~\cite{Carenza:2023old}.

In Fig.~\ref{fig:Snu_Main} we show the predicted IA spectrum for different sterile neutrino masses in the $|l|<60^{\circ}$-$|b|<10^{\circ}$ region, up to a hundred MeV.  We compare these predictions with COMPTEL~\cite{COMPTEL_6010} and EGRET~\cite{Strong_2004} diffuse data, with the former being the most constraining dataset. We stress that one could also obtain constraints from Fermi-LAT at the lowest edge of its energy range, where it's accuracy is lower. We leave this exercise for future work as a possibility to further optimize our analysis.

\begin{figure}[t!]
    \centering
    \includegraphics[width=0.9\columnwidth]{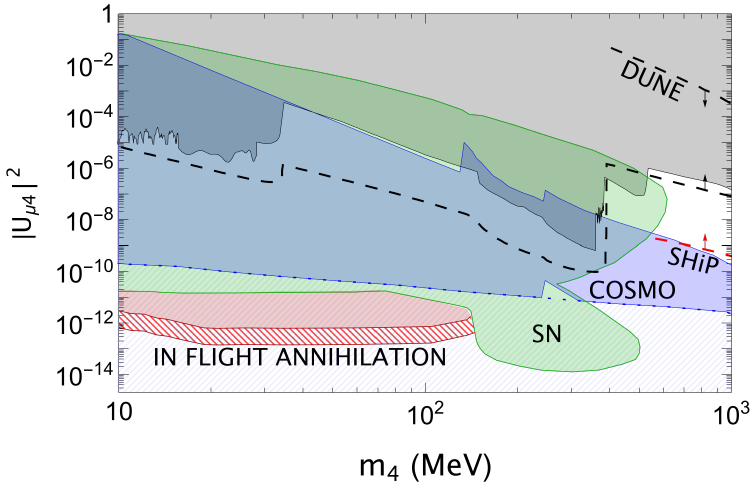}
    \includegraphics[width=0.9\columnwidth]{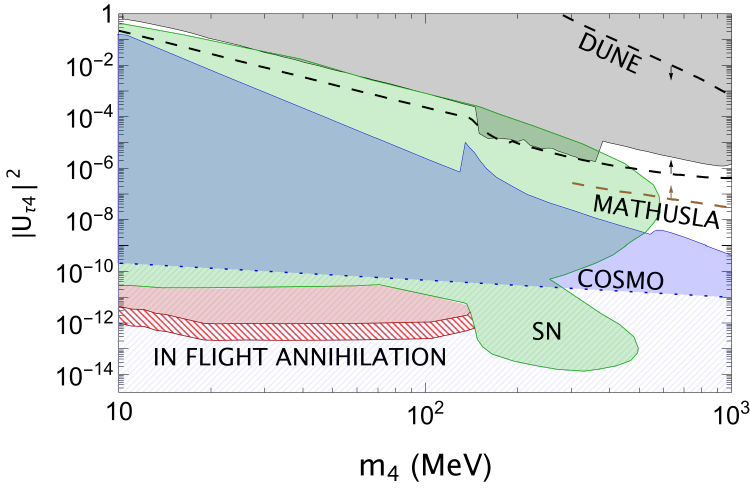}
    \caption{Overview of the astrophysical SN bounds~\cite{Carenza:2023old} (green region), cosmology~\cite{Sabti:2020yrt,Boyarsky:2020dzc,Mastrototaro:2021wzl} (blue region and hatched area for the out of equilibrium extension) and laboratory experiments~\cite{Bolton:2019pcu} (gray region) for sterile neutrinos mixed with $\nu_{\mu}$ (upper panel) or $\nu_\tau$ (lower panel). The dashed lines represent the sensitivities of the future experiments DUNE~\cite{Ballett:2019bgd,Krasnov:2019kdc} (black) and MATHUSLA~\cite{Chou:2016lxi} (brown). The IA bound discussed in this work (red region, with hatched area corresponding to the uncertainty) significantly probes a sizable portion of the parameter space for sterile neutrinos with small mixing angle.}
    \label{fig:comprensive_cons}
\end{figure}

Remarkably, we obtain the most stringent bound on MeV-scale sterile neutrinos,  more than an order of magnitude stronger than laboratory~\cite{Bolton:2019pcu} and astrophysical SN bounds~\cite{Carenza:2023old}, as shown in red in Fig.~\ref{fig:comprensive_cons}. The hatched band corresponds to systematic uncertainties, which mainly consist of uncertainties on the electron density and distribution, the gas phase and temperature where positrons thermalize and the propagation of the positrons. 
To estimate the uncertainties in the predicted positron flux related to our propagation setup, we create two extreme and opposite scenarios, a pessimistic and optimistic one, that lead to the hatched region shown in Fig.~\ref{fig:comprensive_cons}. 

These scenarios mainly differ from our benchmark setup in the normalization of the diffusion coefficient characterizing the diffusion process (i.e. $D_0$), the volume where CRs are confined in the Galaxy (through the value of the halo height $H$) and the level of reacceleration, parameterized in the Àlfven speed $V_A$~\cite{1995ApJ...441..209H, 1998ApJ...509..212S}. In the benchmark case, we use $D_0 = 5.2\times10^{28}$~cm$^2$/s, $H=8$~kpc and
$V_A=13.4$~km/s (see Refs.~\cite{DelaTorreLuque:2023huu,  DelaTorreLuque:2023olp, seo1994stochastic} for more details).
The pessimistic setup features the scenario with very slow diffusion, where we adopt a very small value of the halo height, compared to typical values obtained in CR analyses, $H=3$~kpc. We keep the ratio $H$/$D_0$ to the one that reproduces CR secondary ratios at Earth (i.e. the same ratio as with our benchmark propagation parameters). In addition, we assume no reacceleration (i.e. $V_A = 0$~km/s), thus leading to very conservative estimations of the positron flux produced from these particles. In the optimistic case, we set  $V_A = 40$~km/s and $H=16$~kpc. We notice that even using a very conservative estimation of the  uncertainty associated with propagation, the limits change by no more than a factor of $4$-$5$ in the whole sterile neutrino mass range.

We stress that only the free-streaming regime, i.e. low mixing angle $U_{\alpha4}<\mathcal{O}(10^{-8})$, was considered because of the existing bounds closing most of the parameter space in the large mixing parameter range. Moreover, in this portion of parameter space, the SN evolution might be significantly altered by strongly coupled sterile neutrinos.

The IA probes a portion of the parameter space constrained with cosmological bounds, which strongly rely on modelling assumptions~\cite{Sabti:2020yrt,Boyarsky:2020dzc,Mastrototaro:2021wzl}. For even smaller couplings, it is possible to extend these constraints but they are strongly dependent on the cosmological assumptions~\cite{Ovchynnikov:2021zyo}. 
In this context, our astrophysical bound relies on different assumptions compared to cosmological ones and it is therefore a valuable tool to independently probe this portion of parameter space for such weakly coupled sterile neutrinos.

\emph{Discussion and conclusion---} In this work, we have presented a novel observable, $\gamma$ ray emission from IA, which is able to provide very strong constraints on the positron production from any exotic source of high energy positrons. We investigate the case of electrophilic FIPs produced in SNe, exemplifying the potential of IA emission to provide constraints from diffuse $\gamma$ ray observations. The case of sterile neutrinos was discussed in detail, and we expect similarly powerful results in the case of dark photons and axion-like particles. In addition, this observable can also be used to constrain the fraction of DM in the form of PBHs or the annihilation or decay rate of DM producing electron-positron pairs as a final state, as we will explore in a separate paper.

We remark that our evaluation should be taken as conservative, given that we are only accounting for direct annihilation of positrons with electrons, as pointed out in the main text. In addition, we did not add the internal bremsstrahlung emission~\cite{BeacomIB}, because it is subdominant with respect to the IA emission, but that could improve the constraints by $\mathcal{O}(10\%)$. 

In conclusion, we demonstrate that this novel $\gamma$ ray emission mechanism can provide very strong constraints in a variety of models for positron emitters, regardless of the specific model, being thus a very powerful tool to probe new physics.

\emph{Acknowledgements---} We would like to thank John Ellis for helpful discussions at the beginning of this work and Tim Linden for feedback on the manuscript. This publication is based upon work from COST Action COSMIC WISPers CA21106, supported by COST (European Cooperation in Science and Technology).
SB is supported by the STFC under grant ST/X000753/1.
The work of PC is supported by the Swedish Research Council (VR) under grants  2018-03641 and 2019-02337.
PDL is supported by the Juan de la Cierva JDC2022-048916-I grant, funded by MCIU/AEI/10.13039/501100011033 European Union "NextGenerationEU"/PRTR. The work of PDL is also supported by the grants PID2021-125331NB-I00 and CEX2020-001007-S, both funded by MCIN/AEI/10.13039/501100011033 and by ``ERDF A way of making Europe''. PDL also acknowledges the MultiDark Network, ref. RED2022-134411-T. This project used computing resources from the Swedish National Infrastructure for Computing (SNIC) under project Nos. 2021/3-42, 2021/6-326, 2021-1-24 and 2022/3-27 partially funded by the Swedish Research Council through grant no. 2018-05973.
The work of LM is supported by the Italian Istituto Nazionale di Fisica Nucleare (INFN) through the ``QGSKY'' project and by Ministero dell'Universit\`a e Ricerca (MUR).
\bibliographystyle{apsrev4-1}
\bibliography{references.bib}

%\clearpage

\end{document}